\documentstyle[preprint,prl,aps]{revtex}
\begin{document}
\draft
\title{Nonlinear Conductance for the Two Channel Anderson Model}
\author{Matthias H. Hettler,$^1$ Johann Kroha,$^2$ and Selman Hershfield$^1$}
\address{$^1$Department of Physics and National High Magnetic Field Laboratory,
University of Florida, Gainesville, FL 32611}
\address{$^2$Laboratory of Atomic and Solid State Physics,
Cornell University, Ithaca, NY 14853}
\date{May 20, 1994}
\maketitle
\begin{abstract}
Using the integral equations of the Noncrossing Approximation,
the differential conductance is computed as a function of voltage
for scattering from a two channel Kondo impurity in a point contact.
The results compare well to experimental data on Cu point contacts
by Ralph and Buhrman.  They support a recently proposed
scaling hypothesis, and also show finite temperature corrections
to scaling in agreement with experiment.
The conductance signal is predicted to be asymmetric
in the bias when the impurity is not equally coupled to left and right
moving electrons.
\end{abstract}
\pacs{PACS numbers: 72.10-d, 72.15.Fk, 72.10.Qm, 63.50.+x}

\narrowtext
The two channel Kondo model\cite{nozbl},
or equivalently the Kondo limit of the
two channel Anderson model \cite{schriwo},
has been applied to a wide variety of
interesting physical systems:
heavy fermion compounds\cite{cox2,sea},
high $T_c$ superconductors \cite{cox1},
and two level systems (TLS) in
metals\cite{zv1,mur,rabu,rlvdb}.
Although this model contains many of the salient features
of the experiments, e.g. marginal-Fermi-liquid
behavior \cite{varma,cox1},
the experimental proof for the physical existence of two
channel Kondo impurities is far from certain.

One of the strongest experimental cases for the existence of
two channel Kondo behavior is an experiment
by Ralph and Buhrman \cite{rabu} on clean Cu point contacts.
At low temperatures their samples show the correct temperature,
magnetic field, and voltage dependence for a two channel Kondo
impurity such as a TLS with electron assisted tunneling.
Recently\cite{rlvdb}, the data has also been shown to be consistent with
a scaling ansatz motivated by the equilibrium Conformal Field Theory (CFT)
solution of the problem\cite{afflud}.
In this paper we compute the differential conductance
for a two channel Kondo impurity
in a point contact and
find excellent quantitative agreement with the above experiment,
lending strong evidence for the existence of two channel Kondo impurities
in this system.

The two channel Kondo model consists of two kinds or channels
of mutually noninteracting conduction electrons which are
coupled to the same impurity spin via an exchange interaction\cite{nozbl}.
For the physical realization  of interest here \cite{rabu} a TLS plays
the role of the impurity spin. The states of the TLS correspond to the impurity
spin-up and spin-down states. The electrons scattering from the
impurity may be characterized by parity. The different parity states
play the role of spin-up and spin-down electrons in the conventional Kondo
problem. Thus, parity plays the role of the {\it active}
degree of freedom altered by the interaction, while the
physical spin is a {\it spectator} degree of freedom, which is
conserved by the interaction and which constitutes the two
channels of the model \cite{zv1}.

It is convenient  to represent this system by an Anderson  hamiltonian
of a particle on an impurity level far below the Fermi surface
hybridizing with the two channels of conduction electrons.
We compute the differential conductance within the
Non Crossing Approximation (NCA) for the infinite U  Anderson model
in the Kondo limit\cite{cole,muhart,bickers}.
The NCA has been very successful in describing the one channel Kondo problem
except for the appearance of spurious nonanalytic behavior at a
temperature far below the Kondo temperature $T_K$.
These spurious low-T properties are due to the fact that the  NCA neglects
vertex
corrections responsible for restoring the low temperature Fermi liquid
behavior of the one channel model\cite{grewe}.
However, it has recently been shown \cite{cox3} that for the two
channel problem, where the
complications of the appearance of a Fermi liquid fixed point are
not present, the NCA does give the exact low-frequency power law behavior of
the impurity spectral function $A_d(\omega)$ at zero
temperature. Therefore, we expect to achieve a correct description
for quantities involving $A_d$ (like the conductance).

In the NCA approach the electron on the impurity is represented in terms of
a pseudo Fermion operator, $f$, and a slave Boson operator, $b$ \cite{barnes}.
In order to calculate the conductance at finite bias, the NCA must be
generalized using nonequilibrium Green functions
\cite{langreth}. One solves for both
the retarded Green functions for these operators,
$G_r^f$ and $G_r^b$, and for the `lesser' Green functions
$G^{f}_{<}$ and $G^{b}_{<}$,
which contain information about the nonequilibrium distribution function.
The integral equations for these four functions have been
solved for the one channel case by
Meir, Wingreen, and Lee \cite{mewin}.
However, our numerical implementation \cite{kroha} is different from theirs,
and we are able to
go more than two orders of magnitude lower in temperature,
deep into the low temperature scaling regime described below.

In the Anderson model
the characteristic energy scale,
the Kondo temperature, $T_K$, depends on the width, $\Gamma$,
and position, $E_d \ll -\Gamma$, of the bare Anderson impurity level
\cite{muhart}.
Besides the voltage, $V$, and temperature, $T$, we can also vary
the couplings of the impurity to left and right moving electrons
$\Gamma _L$ and $\Gamma _R$, normalized
to the width $\Gamma $ ($\Gamma _L + \Gamma _R=1 $).
Except for
Fig. 4 all the numerical results have $\Gamma _L = \Gamma _R$.

Letting $F_{eff}(\omega ) = \Gamma _L F(\omega +eV/2)
+ \Gamma _R F(\omega -eV/2)$, where $F(\omega )=1/(1+e^{\beta \omega})$,
and using the conventions of
M\" uller--Hartmann \cite{muhart} for the spectral functions,
$A(\omega )= -ImG_r^f(\omega )/\pi$,
$B(\omega )= -ImG_r^b(\omega )/\pi$,
and the `lesser' Green functions,
$a(\omega )= G^{f}_{<}(\omega )/2\pi$ and
$b(\omega )= G^{b}_{<}(\omega )/2\pi$,
the nonequilbrium NCA equations for the $N=2$ spin, $M=2$ channel
Anderson model are \cite{dos}
\begin{eqnarray}
\frac{B(\omega)}{|G^{b}_{r}(\omega)|^{2}} &=&
   \Gamma N \int \frac{d\epsilon}{\pi}\,
   A(\omega + \epsilon) F_{eff}(\epsilon ) \\
\frac{A(\omega)}{|G^{f}_{r}(\omega)|^{2}} &=&
   \Gamma M \int \frac{d\epsilon}{\pi}\,
   B(\omega - \epsilon) \left[ 1-F_{eff}(\epsilon )\right] \\
\frac{b(\omega)}{|G^{b}_{r}(\omega)|^{2}} &=&
   \Gamma N \int \frac{d\epsilon}{\pi}\,
   a(\omega + \epsilon) \left[ 1-F_{eff}(\epsilon) \right]  \\
\frac{a(\omega)}{|G^{f}_{r}(\omega)|^{2}} &=&
   \Gamma M \int \frac{d\epsilon}{\pi}\,
   b(\omega -\epsilon) F_{eff}(\epsilon )
\end{eqnarray}
The true impurity spectral function, $A_d$,
is computed from the slave Green functions via
the convolution
\begin{eqnarray}
 A_{d}(\omega) = \int \frac{d\epsilon}{\pi}
\left[\, a(\epsilon) B(\epsilon -\omega) + A(\epsilon) b(\epsilon -\omega)
\right] .
\end{eqnarray}

A point contact consists of two leads joined by a small constriction.
Any additional scattering in the vicinity of the constriction should
cause a decrease in the conductance because it impedes the flow of electrons.
On the other hand, in a tunnel junction, tunneling may be assisted by
electrons hopping on impurities in the junction, increasing the
conductance.  Thus, it is not surprising that when we generalize earlier
calculations for the nonlinear current through
a tunnel junction \cite{sdw,mewin2} to the case of a point contact,
we find a similar expression for the current,
except for an overall minus sign:
\begin{eqnarray}
I-I_0\propto -\int d\omega A_{d}(\omega)[F(\omega -V/2)-F(\omega +V/2)],
\label{curr}
\end{eqnarray}
where $I$ and $I_0$ are the currents with and without the impurity.
In deriving Eq.\ \ref{curr} we have assumed that the point contact is clean,
namely the transmission coefficients are close to unity, and that
all hopping matrix elements are slowly varying on the scale of
the Kondo temperature.
The conductance, $G(V,T)$, is computed by taking a numerical derivative,
$dI/dV$.  We will assume that the background conductance $dI_0/dV$ is ohmic.

In Figure 1 we show the zero bias conductance, $G(0,T)$, computed in
this manner ($\Gamma _L = \Gamma _R$).
As expected \cite{afflud}, the conductance shows a $T^{1/2}$
dependence at low temperatures with deviations starting at about
1/4 $T_K$.  The Kondo temperature is determined by the width at
half maximum of the zero bias impurity spectral function, $A_d$,
at the lowest calculated temperatures (see inset).
The slope of the $T^{1/2}$ behavior defines the constant
$B_\Sigma$:
\begin{equation}
G(0,T) - G(0,0) =  B_{\Sigma} T^{1/2} .
\end{equation}
The experimental data
also show a $T^{1/2}$- dependence, but it is
difficult to deduce an accurate estimate of $T_{K}$ by looking at the
deviations from $T^{1/2}$ behavior (assuming that they would occur at
1/4 $T_K$).
An educated guess gives $T_{K}$ to be about 8 Kelvin for sample 1 and 2
and significantly less for sample 3 of Ref.\ \cite{rlvdb}.

Recently, it has been proposed from a CFT solution of the problem in
equilibrium that the experimental data show
scaling of the conductance $G$ as a function of voltage bias $V$
and temperature $T$ of the form \cite{rlvdb}
\begin{equation}
G(V,T) - G(0,T) =  B_{\Sigma} T^{1/2} H( (A\frac{eV}{k_{B}T})), \label{scal}
\end{equation}
where $H$ is a universal scaling function
($H(0) = 0$ and $H(x) \sim x^{1/2}$ for $x >>$1) and $B_{\Sigma}$ and $A$
are nonuniversal constants.
In order to examine whether this ansatz is correct,
in Fig. 2 the rescaled conductance is plotted as a function of
$(eV/k_{B}T)^{1/2}$ for the numerical data (\,a\,) and
the experimental data (\,b, for the best sample (1)).
Motivated by the symmetry of the experimental conductance-voltage curves
we choose equal coupling to left and right moving electrons,
$\Gamma _L = \Gamma _R$ (see also Fig. 4).
Considering that after fixing $B_{\Sigma}$ using Eq. (7)
there are no adjustable parameters,
the agreement is extraordinary.
The collapse of the various $T$ curves
at low bias and the linear behavior for
the low T curves in the range of $2 < (eV/k_{B}T)^{1/2} <4$
is in agreement with the proposed scaling ansatz Eq.\ \ref{scal}.
However, the slope of the
linear part shows temperature dependence for both the
experimental and the numerical data.
This is not contradictory to the scaling ansatz, but it
does show that there are significant temperature dependent
corrections to scaling within the present approach.

To analyze the scaling plots in more detail,
the low bias portion is replotted in
Fig. 3(a).
The  conductance  follows an approximate $(eV/k_{B}T)^{2}$- behavior
even for $(eV/k_{B}T) > 1$
before it levels off and enters the $(eV/k_{B}T)^{1/2}$- region at higher
bias.
The prefactor of the quadratic dependence
shows no observable temperature dependence
until approximately $0.1 T_K$ and consequently
obeys the scaling ansatz.

In Fig. 3(b) the slopes of the straight line fits of the linear regions
in Fig. 2 are plotted as a function of temperature.
Both the numerical data and the experiment show clear
temperature dependence.
Although $B_\Sigma$ may be determined directly from
the zero bias conductance, the Kondo temperature is
more difficult to determine experimentally.  In Fig. 3(b) we have
chosen values for the Kondo temperature which are consistent
with the estimates from the deviation of the zero bias conductance
from $T^{1/2}$ behavior.
The resulting curves are in good quantitative agreement.
In order to show that the
experimental and numerical curves indeed coincide (for a given
$T/T_K$) we also compare the intercepts of the straight line
fits  for the same Kondo temperatures (inset Fig. 3 (b)).
The numerical data do fall right in the middle of the scatter
from the three different samples \cite{fit}.
Note that our theory does not have the additional parameter, $A$,
of Eq.\ \ref{scal}, but adjusting the Kondo temperature has some
of the same effect as adjusting $A$.

Finally, we have also studied the conductance in cases where the impurity is
asymmetrically coupled to left and right moving electrons.
The NCA equations have no symmetry
for $\Gamma_{L} \leftrightarrow \Gamma_{R}$; however, they are symmetric
under the parity operation $\Gamma_{L} \leftrightarrow \Gamma_{R}$ and
$V \leftrightarrow - V$. Figure 4 shows the conductance signal one
should observe in an asymmetrically coupled sample. The conductance
is clearly a nonuniversal function of the couplings $\Gamma_{L}$
and $\Gamma_{R}$. The experimental
data are symmetric to within experimental error,
indicating that left and right moving electrons are equally
coupled to the TLS.
We think this is reasonable for a point contact where the
impurity has no particular asymmetry with respect to left and
right moving electrons, but not for a tunnel
junction, where one would expect exponential dependence of the couplings
on the location of the impurity.

In conclusion, we have performed numerical evaluations of the
NCA integral equations for the two channel Anderson model
out of equilibrium. We find outstanding agreement with
data from an experiment on Cu point
contacts \cite{rlvdb}.
Scaling of the conductance at low bias ($eV<k_B T$) and temperatures is
verified. As $V$ and $T$
are increased, the calculation exhibits finite-$T$ corrections
to scaling, again in agreement with experiment.
Asymmetric conductance signals
are predicted for impurities asymmetrically coupled to the leads.
Thus, this work lends strong support to the existence of two channel Kondo
impurities in the Cu point contacts of Ralph and Buhrman.

The authors have benefited from discussions with
D. Ralph, V. Ambegaokar, R. Buhrman, J. von Delft, L. Borkowski,
P. Hirschfeld, K. Ingersent, A. Ludwig, and A. Schiller.
This work was supported by NSF grant DMR9357474, the U. F.
D.S.R., and the NHMFL (M.H. and S.H.), and by a
Feodor Lynen Fellowship of the Alexander v. Humboldt Found. (J.K).


\begin{figure}
\caption{Temperature dependence of the zero bias conductance
($\Gamma _L = \Gamma _R$).
The zero bias conductance has $T^{1/2}$- dependence for $T< T_K/4$.
This can be used to
roughly extract $T_{K}$ from the experimental data.
Inset: The impurity spectral function $A_{d}(\omega)$ for several voltages.
The width at half maximum of the zero bias spectral function (dotted curve)
determines $T_K$.
As the voltage is increased to $eV=k_{B}T_{K}$ the Kondo resonance
is reduced (solid curve).  At very large bias the resonance
shows a shoulder and eventually two peaks.  In this paper we compare
theory and experiment in the scaling regime, $T\ll T_K$.
}
\label{fig1}
\end{figure}
\begin{figure}
\caption{Scaling plots of the conductance for (a) theory and (b)
experiment [13].
With $\Gamma _L = \Gamma _R$ and
$B_\Sigma$ determined from the zero bias conductance
(see Fig. 1), there are no adjustable parameters.
There are roughly two regimes in these plots.
For $(eV/k_BT)^{1/2} < 1.5$ the curves collapse onto a single
curve and the rescaled conductance is proportional to
$(eV/k_BT)^2$.
For $2<(eV/k_BT)^{1/2} <4$ the rescaled conductance
is linear on these plots. There are substantial
corrections to scaling even at temperatures small compared to $T_K$
(see figure 3 (b)).
At even larger biases this linear behavior rounds off, indicating
the breakdown of scaling.
The temperatures in the theory and experimental plots are
in units of $T_K$ and Kelvin, respectively.}
\label{fig2}
\end{figure}
\begin{figure}
\caption{Quantitative analysis of the scaling plots at (a) low bias
and (b) high bias.
(a) The low bias rescaled conductance as a function of $(eV/k_BT)^2$
(curves are offset).
Both theory and experiment (sample 1 of Ref. [9])
show quadratic behavior at low bias.
The symbols correspond to the temperatures shown in Fig. 2 .\,
(b) The slope of the straight line fits
of the linear region in Fig. 2  as a function
of temperature for sample 1 ($\circ $,+), sample 2 ($\Diamond,\Box $),
sample 3 ($\times,\ast $) (for $V > 0,\,V <0$)
and the numerical data ($\triangle $).
The Kondo temperature for the experimental curves
is consistent with  the deviation of the
zero bias conductance from $T^{1/2}$- behavior.
All curves drop with increasing temperature even at temperatures small
compared to the Kondo temperature. The inset shows the behavior of the
intercepts of the fits.}
\label{fig3}
\end{figure}
\begin{figure}
\caption{Dependence of the conductance on the coupling of the impurity
to left (right) moving electrons, $\Gamma _{L(R)}$,
with $\Gamma _L +\Gamma _R =1$.
In Figs. 1-3 $\Gamma _L$ equals $\Gamma_R$.
For a typical low temperature ($T = 0.01 T_{K}$) one observes an asymmetric
signal, $G(V,T)\ne G(-V,T)$ for $\Gamma_{L} \neq \Gamma_{R}$.
Note that $B_\Sigma$, which depends on $\Gamma _L$, has been divided out.
These curves show that the scaled conductance is not universal
if one allows for asymmetric coupling of the impurity to left
and right moving electrons.}
\label{fig4}
\end{figure}

\begin{references}
\bibitem{dos} Instead of the flat density of states (DOS) of Eqs. (1-4),
              we use a gaussian DOS, which can be included in $F_{eff}$.
\bibitem{fit} We can also directly identify the experimental curves
with the numerical data over the scaling range.
\end{references}
\end{document}